\documentclass[letterpaper,11pt,reqno]{amsart}

\makeatletter
\usepackage{amssymb}
\usepackage{latexsym}
\usepackage{amsbsy}
\usepackage{amsfonts}
\usepackage{caption}
\usepackage{subcaption}
\usepackage{hyperref}
\usepackage{graphicx}
\usepackage{enumerate}
\usepackage{enumitem}
\usepackage{mathtools}
\usepackage{color}

\usepackage{tikz}

\def\marginpar#1{\ignorespaces}

\DeclareMathOperator\Dir{Dir}
\DeclareMathOperator\var{Var}

\newtheorem{theorem}{Theorem}[section]

\newtheorem{proposition}[theorem]{Proposition}
\newtheorem{corollary}[theorem]{Corollary}

\newtheorem{assump}[theorem]{Assumption}

\numberwithin{equation}{section}
\makeatother
\begin{document}
\title[Proof of Stake]{Trading and wealth evolution in the Proof of Stake protocol}

\author[Wenpin Tang]{{Wenpin} Tang}
\address{Department of Industrial Engineering and Operations Research, Columbia University. 
} \email{wt2319@columbia.edu}

\date{\today} 
\begin{abstract}
With the increasing adoption of the Proof of Stake (PoS) blockchain,
it is timely to study the economy created by such blockchain.
In this chapter, we will survey recent progress on the trading and wealth evolution in a cryptocurrency
where the new coins are issued according to the PoS protocol.
We first consider the wealth evolution in the PoS protocol assuming no trading, and focus on the problem of decentralisation. 
Next we consider each miner's trading incentive and strategy through the lens of optimal control,
where the miner needs to trade off PoS mining and trading.
Finally, we study the collective behavior of the miners in a PoS trading environment by a mean field model.
We use both stochastic and analytic tools in our study.
A list of open problems are also presented.
\end{abstract}

\maketitle


\section{Introduction}
\label{sc1}

\quad A blockchain is a digit ledger allowing the secure transfer of assets
in a distributed network without an intermediary, 
hence achieving decentralisation.
As the internet is a technology to facilitate the digit flow of information,
the blockchain is a technology to facilitate the digit exchange of value. 
In the past decade, 
blockchain technology has advanced tremendously,
with a wide range of applications including
cryptocurrency \cite{Naka08, Wood14},
healthcare \cite{MC19, TPE20},
supply chain \cite{CTT20, SK19},
and non-fungible tokens \cite{Dow22, WL21}.
See Part 2 of this book for other applications of the blockchain. 
Recently, a large number of financial institutions 
seek to launch crypto exchanges in the stock market \cite{MN23}. 

\quad At the core of a blockchain is the consensus protocol, 
which specifies a set of voting rules for the participants (miners or validators) 
to agree on an ever-growing log of transactions
so as to form a distributed ledger.
There are several existing blockchain protocols, 
among which the most popular are {\em Proof of Work} (PoW, \cite{Naka08})
and {\em Proof of Stake} (PoS, \cite{KN12, Wood14}):
\begin{itemize}[itemsep = 3 pt]
\item
In the PoW protocol, miners compete with each other by solving a hashing puzzle.
The miner who solves the puzzle first receives a reward (a number of coins) 
and whose work validates a new block’s addition to the blockchain. 
Hence, while the competition is open to everyone, 
the chance of winning is proportional to a miner’s computing power.
The PoW coins include Bitcoin and Dogecoin.
\item
In the PoS protocol, there is a bidding mechanism to select a miner to do the work of
validating a new block.
Participants who choose to join the bidding are required to commit some stakes (coins they own),
and the winning probability is proportional to the number of stakes committed.
The PoS coins include Ethereum and BNB.
\end{itemize}

\quad As of July 15, 2023, \texttt{Cryptoslate} lists 326 PoW coins with a total \$628B (51\%) market capitalisation,
and 248 PoS coins with a total \$321B (26\%) market capitalisation. 
One major pitfall of the PoW protocol is that competition among the miners has led to exploding levels of energy consumption,
and hence raised the issue of sustainability \cite{Mora18, PS21}.
\cite{CK22, CHL20, Saleh19} also discussed the drawbacks of the PoW blockchain from economic perspectives. 
These concerns have created a strong incentive among blockchain practitioners
to switch from the PoW to the PoS ecosystem,
as was pioneered by Ethereum 2.0 in September 2022 \cite{Wick21}.

\begin{figure}[h]
\centering
\includegraphics[width=0.65\columnwidth]{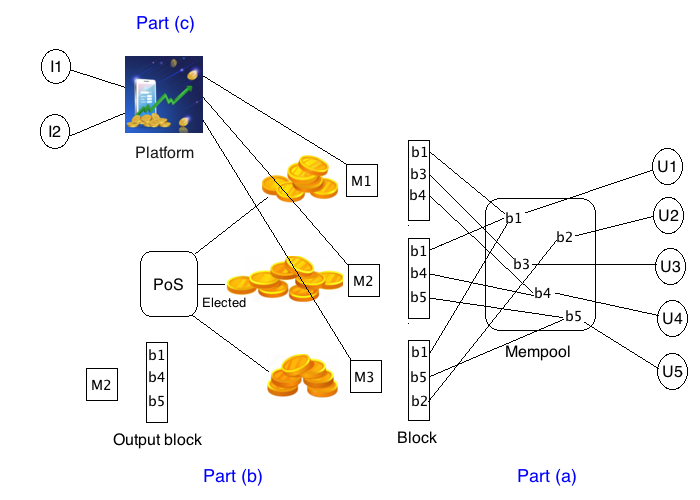}
\caption{Miner-user-investor activities under the PoS protocol.}
\label{fig:0}
\end{figure}

\quad In this chapter,
we present recent research on the PoS protocol,
with focus on its wealth evolution.
There are three major components in the PoS ecosystem:
\begin{enumerate}[itemsep = 3 pt]
\item[(a)]
{\em User-miner interface}.
The users seek to get their transactions settled, 
and published on the blockchain by the miners. 
Since each block has a maximum capacity, 
most blockchains adopt a ``pay your bid" auction,
in which the users bid to have the miners include their transactions in the blockchain.
(In general, the more a user bids,
the more likely her transaction will be settled shortly.)
The activity of the user-miner interface relies on the blockchain adoption,
and the problem is to design a good transaction fee mechanism,
e.g. satisfying some incentive-compatible conditions.
\item[(b)]
{\em Built-in PoS protocol}.
Each miner selects a set of transactions from the mempool (according to the bids mentioned in (a)),
and includes them into a block.
As explained earlier, 
the miners then commit their stakes in a PoS election,
and the elected miner gains the right to add the new block to the blockchain.
In return, the elected miner will receive transaction fees from the users,
and block rewards from the blockchain.
The PoS protocol may have additional hard-coded rules, e.g. the longest chain.
The key issue is the security level facing to various attacks. 
\item[(c)] 
{\em Speculation and trading}.
As the blockchain is a digit exchange vehicle, 
there is a cryptocurrency (crypto) attached to it. 
Along with the increasing blockchain adoption, 
crypto has become a new financial instrument. 
This leads to the crypto trading. 
The trading parties are the miners and the investors (i.e. the crypto market).
The investors may be the blockchain users who trade the crypto for use,
or the speculators who seek profit from crypto holdings. 
The problem is to understand the trading strategy and wealth evolution of
the participants in the crypto market.
\end{enumerate}
See Figure \ref{fig:0} for an illustration of the aforementioned components in the PoS ecosystem.
Here we concentrate on part (c). 
Refer to \cite{BN19, CS23, TY23IP} for discussions related to part (a),
and \cite{BD19, DPS19, KR17, Saleh21} for developments in part (b).

\quad As the readers may have observed, 
the miners play a particularly important role in the blockchain ecosystem:
they manage the user-miner interface (in part (a));
they maintain the blockchain (in part (b));
they provide liquidity in the crypto trading (in part (c)).
For the miners,
they can commit their stakes to participate in the PoS mining process,
trade their stakes on the crypto market for instantaneous profit,
or a combination of the two. 
The most obvious questions are
how the miner allocates her stakes between PoS mining and trading (called a {\em strategy}),
and what is her wealth evolution. 
The former question is concerned with the miner's optimal strategy,
while the latter studies the level of decentralisation in the PoS economy. 
Being more specific, we ask:
\begin{enumerate}[itemsep = 3 pt]
\item
Does the PoS protocol lead to centralisation or the rich-get-richer phenomenon
(assuming no trading)?
Under what conditions do the miners prefer not to trade?
\item
For each individual miner, what is her best strategy?
\item
What is the wealth evolution of the whole (miner) population
if each miner follows the best response to the others?
\end{enumerate}

\quad We will answer these questions in the following sections. 
The question (1) hinges on a P\'olya urn representation of the PoS election/protocol. 
To address the question (2), we formulate and solve an optimal control problem.
The optimal control framework also allows us to analyse the question (3) via a mean field model. 
We conclude with a few open problems and future directions. 
We emphasise that each PoS blockchain may have specific rules
(e.g. block validity, the longest chain...etc),
and we will not take these blockchain-specific rules into account.
Our analysis applies to the generic PoS protocol
which will be defined accordingly. 

\section{Stability of the PoS protocol}
\label{sc2}

\quad We consider the question (1) in this section. 
Recall that the miners who choose to join the PoS election are required to commit their coins,
and the winning probability is proportional to the number of coins committed.
Assume in this section that no trading is allowed,
and all the miners will participate in the PoS mining process.
Hence, the number of coins committed by each miner 
is equal to that she owns. 
At first glance, the miner who owns the largest number of coins
is more likely to win the PoS election,
which will in turn generate more and more coins for her. 
This is called the rich-get-richer phenomenon, 
which fundamentally violates the decentralised nature of any blockchain.
We will show that the wealth evolution of a PoS miner 
depends on the reward type and her coin possession level. 

\quad Now we describe formally the PoS protocol without trading. 
Time is discrete, indexed by $t \in \{0,1,\ldots\}$. 
Let $K$ be the number of miners, 
and $N \in \mathbb{R}_{+}$ be the number of initial coins in the PoS blockchain.
The miners are indexed by $[K]: = \{1, \ldots, K\}$,
and miner $k$'s initial coins are $n_{k,0}$ with $\sum_{k = 1}^K n_{k,0} = N$.
We define the {\em share} as the fraction of coins each miner owns. 
So the initial shares $(\pi_{k, 0}, \, k \in [K])$ are given by
\begin{equation}
\label{eq:share0}
\pi_{k, 0}: = \frac{n_{k,0}}{N}, \quad k \in [K].
\end{equation}
Similarly, denote by $n_{k,t}$ the number of coins owned by miner $k$ at time $t$, and the corresponding share is
\begin{equation}
\label{eq:sharet}
\pi_{k,t}:= \frac{n_{k,t}}{N_t}, \quad k \in [K], \quad \mbox{with } N_t:= \sum_{k=1}^K n_{k,t}.
\end{equation}
Here $N_t$ is the total number of coins at time $t$, with $N_0 = N$.
(We shall often refer to $N_t$ as the ``volume of coins", or simply ``volume".)

\quad At time $t$, miner $k$ is selected at random with probability $\pi_{k, t-1}$.
Once selected, the miner receives a deterministic reward of $R_t \in \mathbb{R}_{+}$ coins 
(which may include transaction fees and block rewards).
Denote by $S_{k,t}$ the random event that miner $k$ is selected at time $t$. 
So the number of coins owned by each miner evolves as  
\begin{equation}
\label{eq:TPolya}
n_{k,t} = n_{k, t-1} + R_t 1_{S_{k,t}}, \quad k \in [K].
\end{equation}
Note that the volume satisfies $N_t = N_{t-1} + R_t$. 
Combining \eqref{eq:sharet} and \eqref{eq:TPolya} yields a recursion of the shares:
\begin{equation}
\label{eq:Dshare}
\pi_{k,t} = \frac{N_{t-1}}{N_t} \pi_{k, t-1} + \frac{R_t}{N_t} 1_{S_{k,t}}, \quad k \in [K].
\end{equation}
which is a (time-dependent) P\'olya urn model \cite{Pem07}.

\quad We consider the long-time evolution of the shares $(\pi_{k,t}, \, k \in [K])$.
Let $\mathcal{F}_t$ be the filtration generated by the random events $(S_{k, r}: k \in [K], r \le t)$.
Observe that for each $k \in [K]$,
the process $(\pi_{k,t}, \, t \ge 0)$ is an $\mathcal{F}_t$-martingale.
By the martingale convergence theorem (see \cite[Theorem 4.2.11]{Durrett}),
\begin{equation}
(\pi_{1,t}, \ldots, \pi_{K,t}) \longrightarrow (\pi_{1,\infty}, \ldots, \pi_{K,\infty}) \quad \mbox{as } t \rightarrow \infty \mbox{ with probability }1,
\end{equation}
where $(\pi_{1,\infty}, \ldots, \pi_{K,\infty})$ is some random probability distribution on $[K]$.

\quad To quantify the wealth evolution of miner $k$, there are two obvious metrics:
\begin{equation}
|\pi_{k,t} - \pi_{k,0}| \mbox{ (difference)} \quad \mbox{and} \quad \frac{\pi_{k,t}}{\pi_{k,0}} \mbox{ (ratio)}.
\end{equation}
If $|\pi_{k,t} - \pi_{k,0}|$ is close to $0$, or $\frac{\pi_{k,t}}{\pi_{k,0}}$ is close to $1$ (as $t$ is large),
we say that the share $\pi_{k,t}$ is {\em stable} or {\em concentrated}.
This is the desired case as 
it implies that the PoS protocol will not lead to centralisation. 
Note that if $\pi_{k,t}$ is of constant order,
there is no difference in considering $|\pi_{k,t} - \pi_{k,0}|$ or $\pi_{k,t}/\pi_{k,0}$.
However, when $\pi_{k,t}$ is small,
the two metrics may exhibit different results:
$|\pi_{k,t} - \pi_{k,0}|$ is (trivially) close to $0$ ($0-0$), while $\pi_{k,t}/\pi_{k,0}$ is indeterminate ($0/0$).

\quad First, assume that $R_t \equiv R$ (constant reward),
where the limiting $(\pi_{1, \infty}, \ldots, \pi_{K,\infty})$ can be identified. 
Let $\Gamma(z):=\int_0^{\infty} x^{z-1} e^{-x} dx$ be the Gamma function.
Recall that the Dirichlet distribution with parameters $(a_1, \ldots, a_K)$, 
which we denote by $\Dir(a_1, \ldots, a_K)$, has support on the standard simplex $\{(x_1, \ldots, x_K) \in \mathbb{R}_{+}^K: \sum_{k = 1}^K x_k = 1\}$ and has density:
\begin{equation}
\label{eq:Dirichlet}
f(x_1, \ldots, x_K) = \frac{\Gamma\left(\sum_{k=1}^K a_k \right)}{\prod_{k=1}^K \Gamma(a_k)} \prod_{k=1}^K x_k^{a_k-1}.
\end{equation}
The following theorem elucidates the wealth evolution of a PoS miner with a constant reward. 

\begin{theorem}
\cite{RS21, Tang22}
\label{thm:1}
Assume that the coin reward is $R_t \equiv R > 0$.
Then the miner shares have a limiting distribution
\begin{equation}
\label{eq:limDirichlet}
(\pi_{1,\infty}, \ldots, \pi_{K,\infty}) \stackrel{d}{=} \Dir\left(\frac{n_{1,0}}{R}, \ldots,\frac{n_{K,0}}{R}\right).
\end{equation}
Moreover, 
\begin{enumerate}[itemsep = 3 pt]
\item[(i)]
For $n_{k,0} = f(N)$ such that $f(N) \to \infty$ as $N \to \infty$,
we have for each $\varepsilon > 0$ and for each $t \ge 1$ or $t = \infty$:
\begin{equation}
\label{eq:coninter}
\mathbb{P}(|\pi_{k,t} - \pi_{k,0}| > \varepsilon) \to 0
\quad \mbox{and} \quad
\mathbb{P}\left(\left|\frac{\pi_{k, t}}{\pi_{k,0}} - 1\right| > \varepsilon \right) \to 0,
\quad \mbox{as } N \to \infty.
\end{equation}
\item[(ii)]
For $n_{k,0} = \Theta(1)$, we have for each $\varepsilon > 0$,
$\mathbb{P}(|\pi_{k,\infty} - \pi_{k,0}| > \varepsilon) \to 0$ as $N \to \infty$,
and the convergence in distribution:
\begin{equation}
\label{eq:consmall}
\frac{\pi_{k,\infty}}{\pi_{k,0}} \stackrel{d}{\longrightarrow} \frac{R}{n_{k,0}} \gamma\left( \frac{n_{k,0}}{R}\right), \quad \mbox{as } N \to \infty,
\end{equation}
where $\gamma\left(\frac{n_{k,0}}{R}\right)$ is a Gamma random variable with density $x^{\frac{n_{k,0}}{R}-1} e^{-x} 1_{x > 0}/\Gamma\left(\frac{n_{k,0}}{R}\right)$.
\end{enumerate}
\end{theorem}

\begin{figure}[htb]
  \includegraphics[width=0.45\linewidth]{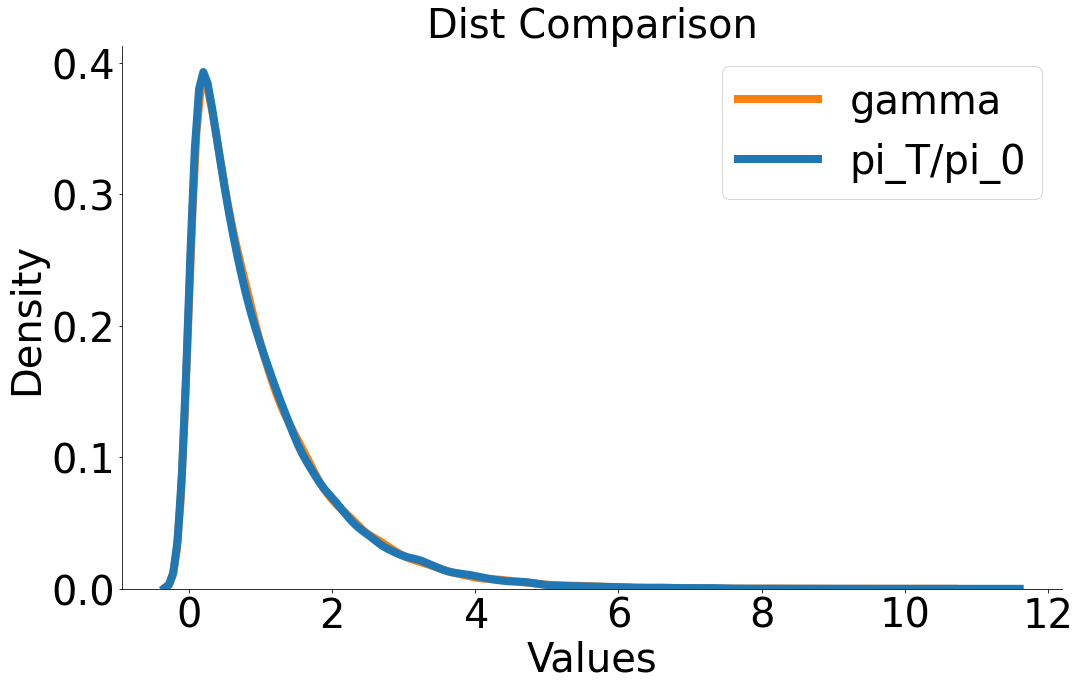}
  \caption{Constant reward: instability of $\pi_{k,t}/\pi_{k,0}$ for small miners. Blue curve: histogram of $\pi_{k,50000}/\pi_{k,0}$ with $n_{k,0} = R = 1$ and $N = 100$. Orange curve: Gamma distribution.}
  \label{fig:2A}
\end{figure}

\quad Let's make a few comments. 
The theorem reveals a {\em phase transition} of shares in the long run between large and small miners. 
Part ($i$) shows that for large miners, 
their shares are stable 
(which holds not only for extremely large miners with initial coins $n_{k,0} = \Theta(N)$
but for less rich large miners with $n_{k,0} \gg 1$, $n_{k,0} = o(N)$.)
On the other hand,
the evolution of shares for small miners has a different limiting behavior. 
Part ($ii$) shows that a small miner's share is volatile in such a way that 
the ratio $\pi_{k,\infty}/\pi_{k,0}$ is close to a gamma distribution independent of the initial coin offerings, 
and hence $\var(\pi_{k,\infty}/\pi_{k,0}) \approx \frac{1}{n_{k,0}}$.
For instance, if $n_{k,0} = R = 1$ the limiting distribution of the ratio $\pi_{k,\infty}/\pi_{k,0}$ reduces to the exponential distribution with parameter $1$. 
(See Figure \ref{fig:2A} for an illustration of this approximation.)
In this case, we have
\begin{equation*}
\mathbb{P}\left(\frac{\pi_{k,\infty}}{\pi_{k,0}} >  \theta \right) \approx e^{-\theta} \quad \mbox{as } N \to \infty.
\end{equation*}
Thus, with probability $e^{-2} \approx 0.135$ a small miner's share will double, and with probability $1 - e^{-0.5} \approx 0.393$ this miner's share will be halved. 

\quad Next we consider the wealth evolution of a PoS miner with a decreasing reward.
Though the limiting $(\pi_{1, \infty}, \ldots, \pi_{K,\infty})$ is not explicit,
we can still characterise a miner's share stability in terms of her coin possession level.

\begin{theorem} \cite{Tang22}
\label{thm:2}
Assume that the coin reward is $R_t$ with $R_t \ge R_{t+1}$ for each $t \ge 0$.
\begin{enumerate}[itemsep = 3 pt]
\item
If $R_t$ is bounded away from $0$, i.e. $\lim_{t \ge 0} R_t = \underline{R} > 0$, then
\begin{enumerate}[itemsep = 3 pt]
\item[(i)]
For $n_{k,0} = f(N)$ such that $f(N) \to \infty$ as $N \to \infty$, we have for each $\varepsilon > 0$ and each $t \ge 1$ or $t = \infty$:
\begin{equation}
\label{eq:decinter1}
\mathbb{P}\left(\left|\frac{\pi_{k, t}}{\pi_{k,0}} - 1\right| > \varepsilon \right) \to 0, \quad \mbox{as } N \to \infty.
\end{equation}
\item[(ii)]
For $n_{k,0} = \Theta(1)$, we have $\var \left(\frac{\pi_{k, \infty}}{\pi_{k,0}}\right) = \Theta(1)$.
Moreover, there is $c > 0$ independent of $N$ such that for $\varepsilon > 0$ sufficiently small:
\begin{equation}
\label{eq:decsmall1}
 \mathbb{P}\left(\left|\frac{\pi_{k, \infty}}{\pi_{k,0}} - 1 \right| > \varepsilon\right) \ge c.
\end{equation}
\end{enumerate}
\item
If $R_t = \Theta(t^{-\alpha})$ for $\alpha > 0$, then for each $\varepsilon > 0$ and each $t \ge 1$ or $t = \infty$:
\begin{equation}
\label{eq:decinter2}
\mathbb{P}\left(\left|\frac{\pi_{k, t}}{\pi_{k,0}} - 1\right| > \varepsilon \right) \to 0, \quad \mbox{as } N \to \infty.
\end{equation}
\end{enumerate}
\end{theorem}

\quad The theorem distinguishes two ways that the reward function decreases,
leading to different phase transition results. 
Part ($1$) assumes that the reward function decreases to a nonzero value.
In this case, the threshold to identify large and small miners is $n_{k,0} = \Theta(1)$, 
which is the same as that of the PoS protocol with a constant reward.
This may not be surprising,
since the underlying dynamics is not much different from the one with a constant reward. 
For large miners, the ratio $\pi_{k, \infty}/\pi_{k,0}$ is close to $1$;
while for small miners there is the {\em anti-concentration} bound \eqref{eq:decsmall1},
indicating that the evolution of a small miner's share is no longer stable, and may be volatile. 
Part (2) considers a fast decreasing reward $R_t = \Theta(t^{-\alpha})$ for $\alpha > 0$. 
In this case, there is no phase transition, and the ratio $\pi_{k, \infty}/\pi_{k,0}$ concentrates at $1$
for every miner. 

\quad To conclude this section, 
we present the results on the wealth evolution of a PoS miner with an increasing reward.

\begin{theorem}
\label{thm:3}
\cite{Tang22}
Assume that the coin reward $R_t = \rho N_{t-1}^{\gamma}$ for some $\rho > 0$ and $\gamma > 0$.
\begin{enumerate}[itemsep = 3 pt]
\item
If $\gamma > 1$, 
then $\pi_{k, \infty} \in \{0,1\}$ almost surely with
\begin{equation}
\label{eq:incextreme}
\mathbb{P}(\pi_{k, \infty} = 1) = \pi_{k,0}, \quad \mathbb{P}(\pi_{k, \infty} = 0) = 1 - \pi_{k,0}
\end{equation}
\item
If $\gamma < 1$, then
\begin{enumerate}[itemsep = 3 pt]
\item[(i)]
For $n_{k,0} = f(N)$ such that $f(N)/N^{\gamma} \to \infty$ as $N \to \infty$, 
we have for each $\varepsilon > 0$ and each $t \ge 1$ or $t = \infty$:
\begin{equation}
\label{eq:incinter}
\mathbb{P}\left(\left|\frac{\pi_{k, t}}{\pi_{k,0}} - 1\right| > \varepsilon \right)
\to 0 \quad \mbox{as } N \to \infty.
\end{equation}
\item[(ii)]
For $n_{k,0} = \Theta(N^{\gamma})$, we have $\var \left(\frac{\pi_{k, \infty}}{\pi_{k,0}}\right) = \Theta(1)$.
Moreover, there exists $c > 0$ independent of $N$ such that for $\varepsilon > 0$ sufficiently small:
\begin{equation}
\label{eq:incsmall}
 \mathbb{P}\left(\left|\frac{\pi_{k, \infty}}{\pi_{k,0}} - 1 \right| > \varepsilon\right) \ge c.
\end{equation}
For $n_{k,0} = o(N^{\gamma})$, we have $\var \left(\frac{\pi_{k, \infty}}{\pi_{k,0}}\right) \to \infty$ as $N \to \infty$.
\end{enumerate}
\end{enumerate}
\end{theorem}

\begin{figure}[htb]
    \centering
\begin{subfigure}{0.33\textwidth}
  \includegraphics[width=\linewidth]{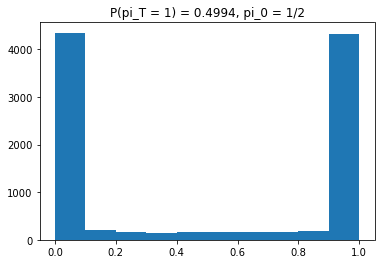}
  \caption{$\pi_{k,0} = \frac{1}{2}$}
\end{subfigure}\hfil
\begin{subfigure}{0.33\textwidth}
  \includegraphics[width=\linewidth]{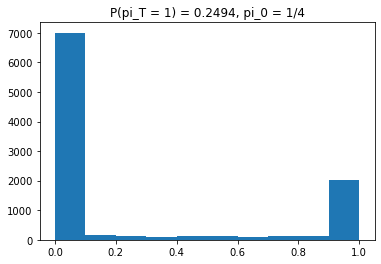}
  \caption{$\pi_{k,0} = \frac{1}{4}$}
\end{subfigure}\hfil
\begin{subfigure}{0.33\textwidth}
  \includegraphics[width=\linewidth]{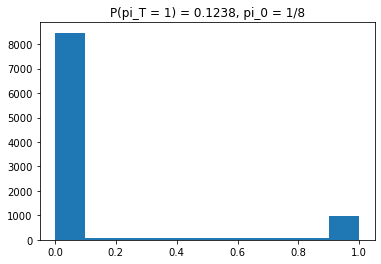}
  \caption{$\pi_{k,0} = \frac{1}{8}$}
\end{subfigure}
\caption{Increasing reward: chaotic centralisation.
Histogram of $\pi_{k, 5000}$ with $\rho = 0.001$, $\gamma = 1.1$, $N = 1000$ and $\pi_{k,0} \in \{1/2, 1/4, 1/8\}$}
\label{fig:9}
\end{figure}

\quad The theorem considers two increasing reward schemes: 
a geometric reward and a sub-geometric one.
Part ($1$) assumes a geometric reward, 
and shows that with probability one, all the shares will eventually go to one miner in such a way that
\begin{equation*}
\mathbb{P}(\pi_k = 1 \mbox{ and } \pi_j = 0 \mbox{ for all } j \ne k) = \pi_{k,0}, \quad k \in [K].
\end{equation*}
We call this {\em chaotic centralisation} 
because the underlying dynamics will lead to the dictatorship, 
with the dictator being selected in a random manner.
(See Figure \ref{fig:9} for an illustration of chaotic centralisation.)
Part ($2$) considers a polynomial reward $R_t = \Theta(t^{\frac{1}{1-\gamma}})$ for $\gamma < 1$.
In this case, there is a phase transition in the stability of $\pi_{k,t}/\pi_{k,0}$, 
with the threshold $n_{k,0} = \Theta(N^\gamma)$.

\quad We also mention that 
it is possible to study the wealth evolution in the PoS protocol
with infinite population ($K = \infty$),
see \cite[Section 3]{Tang22}.

\section{Participation and PoS trading}
\label{sc3}

\quad We consider the question (2) in this section,
and provide conditions under which 
no miner will have incentive to trade (so Theorems \ref{thm:1}-- \ref{thm:3} continue to hold). 
So far, we have not considered the possibility of allowing the miners to 
trade coins (among themselves). 
In the new setting of allowing trading, 
we need to modify the problem formulation presented
in Section \ref{sc2}.
First, for each $k \in [K]$,
let $\nu_{k,t}$ be the number of coins that miner $k$ will trade at time $t$.
Then, instead of \eqref{eq:TPolya}, the number of coins $n_{k,t}$ evolves as
\begin{equation}
\label{eq:TP3}
n_{k, t} = \underbrace{n_{k, t-1} + R_t 1_{S_{k,t}}}_{n'_{k,t}} + \nu_{k,t},
\end{equation}
i.e. $n'_{k,t}$ denotes the number of coins miner $k$ owns in between 
time $t-1$ and $t$, excluding those traded in period $t$.
Note that $\nu_{k,t}$ will be up to miner $k$ to decide, 
as opposed to the random event $S_{k,t}$ which is exogenous; 
in particular, $\nu_{k, t}$ can be negative (as well as positive or zero). 
We will elaborate more on this below, but note that  $\nu_{k,t}$ will be constrained such
that after the updating in (\ref{eq:TP3}) $n_{k, t}$ will remain nonnegative.

\quad Let $\{P_t, \, t \ge 0\}$ be the price process of each (unit of) coin,
 which is a stochastic process assumed to be independent of the 
randomness induced by the PoS selection (specifically, the process $\{S_{k,t}\}$). 
Hence, we augment the filtration $\{\mathcal{F}_t\}_{t \ge 0}$ 
with that of the exogenous price process $\{P_t, \, t \ge 0\}$ to a new filtration denoted
$\{\mathcal{G}_t\}_{t \ge 0}$.
This assumption need not be so far off, as the crypto's price tends to be affected by 
 market shocks 
(such as macroeconomics, geopolitics, breaking news, etc) much more than by trading activities.

\quad Let $b_{k,t}$ denote (units of) the risk-free asset that miner $k$ holds at time $t$, 
and $r_{\tiny \mbox{free}}> 0$ the risk-free (interest) rate.
As we are mainly concerned with the effect of exchanging coins to each individual, 
we only allow miners to trade coins, but not risk-free assets between them.
Hence, each miner has to trade risk-free asset with a third-party instead of trading that with another bidder.

\quad The decision for each miner $k$ at $t$ is a tuple, $(\nu_{k,t}, b_{k,t})$.
Moreover, there is a terminal time, denoted $T_k \ge 1$, 
by which time 
miner $k$ has to sell all assets, including both any risk-free asset and any coins owned at that time.
$T_k$ can either be deterministic or random. 
In the latter case, assume it has a finite expectation,
and is either adapted to $\{\mathcal{G}_t\}_{t \ge 0}$,
or independent of all other randomness (in which case augment $\{\mathcal{G}_t\}$ accordingly).  
We also allow miner $k$ to liquidate prior to $T_k$ at a stopping time $\tau_k$ 
relative to $\{\mathcal{G}_t\}_{t \ge 0}$.
Thus, miner $k$ will also decide at which time $\tau_k$ to stop and exit.
Abuse $\tau_k$ for the minimum of $\tau_k$ and $T_k$.

\quad Let $c_{k,t}$ denote the (free) cash flow  (or, ``consumption'') of miner $k$ at time $t$, i.e.,
\begin{equation*}
 c_{k,t} = (1+r_{\tiny \mbox{free}}) b_{k, t-1} - b_{k, t}  - \nu_{k,t}P_t, 
 \quad \forall 1\le t< \tau_k; \tag{C1}
\end{equation*}
with 
\begin{equation*}
b_{k,0} = 0, \; b_{k,t} \ge 0,\quad
 0 \le n_{k,t} = n'_{k,t} + \nu_{k,t} \le N_t, \quad \forall 1\le t< \tau_k; \tag{C2}
\end{equation*} 
and
\begin{equation*}
c_{k, \tau_k} = (1+ r_{\tiny \mbox{free}}) b_{k, \tau_k- 1} + n'_{k, \tau_k} P_{\tau_k}, 
 \quad \mbox{ and } \nu_{k,\tau_k}=b_{k, \tau_k} = 0.  \tag{C3} 
\end{equation*}

The equation in (C1) is a budget constraint,
which defines what's available for ``consumption'' 
in period $t$.
The requirements in (C2) are all in the spirit of disallowing shorting, 
on both the free asset $b_{k,t}$ and 
the traded coins $\nu_{k,t}$. 
In particular the latter is constrained such that 
$\nu_{k,t} \ge -n'_{k,t}$,
i.e. miner $k$ cannot sell more
than what's in possession at $t$; it also ensures that 
no miner can own a number of coins beyond the total volume ($n_{k,t} \le N_t$).
(C3) specifies how the assets are liquidated at the exit time $\tau_k$: 
both $\nu_{k,\tau_k}$ and $b_{k, \tau_k}$ will be set at zero, 
and all remaining coins $n'_{k, \tau_k}$ liquidated (cashed out at $P_{\tau_k}$ per unit).

\quad Denote by $\tau_k$ and $(\nu,b):=\{(\nu_{k,t}, b_{k,t}), \, 1\le t\le \tau_k\}$
miner $k$'s decision (process) or ``strategy''.
The objective of miner $k$ is to solve the consumption-investment problem:
\begin{equation}
\label{eq:OPT}
U^*_k := \max_{\tau_k, (\nu ,b)}  U_k :=
\max_{\tau_k, (\nu ,b)}  \mathbb{E} \left(\sum_{t=1}^{\tau_k} \delta_k^{t} c_{k,t} \right), 
\quad \text{subject to (C1), (C2), (C3)} ;
\end{equation}
where $\delta_k \in (0,1]$ is a discount factor, a given parameter measuring the risk sensitivity of miner $k$.

\quad We need to introduce two more processes 
that are related and central to understanding the PoS protocol in the presence of trading.  
The first one is $\{M_{t}, \, t\ge 1\}$, where
$M_t := N_t P_t $
denotes the market value of the coins at time $t$.
The second one is  $\{\Pi_{k,t}, \, t\ge 0\}$, for each bidder $k$, defined as follows:
\begin{equation}
\label{eq:Pi}
\Pi_{k,0} := n_{k,0} P_0, \quad \mbox{and} \quad 
\Pi_{k,t} := \delta^t_k n'_{k,t} P_t - \sum_{j=1}^{t-1} \delta_k^j \nu_{k, j} P_j , \quad t \ge 1;
\end{equation}
where $n'_{k,t+1}$ follows (\ref{eq:TP3}).
The process $\{\Pi_{k,t}\}$ connects to the utility $U_k$ in \eqref{eq:OPT}.
To see this, summing up both sides of (C1) and (C3) over $t$ (along with $b_{k,0} = 0$ in (C2)), 
we get
\begin{equation}
\label{eq:ut1}
\sum_{t \le \tau_k} \delta^{t}_k c_{k,t} = \sum_{t \le \tau_k} \delta^{t}_k c_{k,t} = \delta_k^{\tau_k} n'_{\tau_k} P_{\tau_k}
-\sum_{t =1}^{\tau_k-1} \delta^{t}_k \nu_{k,t} P_t
+\sum_{t =1}^{\tau_k-1}\delta_k^t \left[(1+r_{\tiny \mbox{free}}) \delta_k - 1\right] b_{k,t}. 
\end{equation}
Observe that the first two terms on the right hand side are equal to $\Pi_{k,\tau_k}$,  so we can rewrite the above as
follows, emphasizing the exit time $\tau_k$ and the strategy $(\nu,b)$, 
\begin{equation}
\label{eq:uk}
U_k (\tau_k, \nu,b) 
=\mathbb{E} \left[ \Pi_{k,\tau_k} (\nu)\right] + \mathbb{E}\left( \sum_{t =1}^{\tau_k -1}\delta_k^t \left[(1+r_{\tiny \mbox{free}}) \delta_k - 1\right] b_{k,t}\right);  
\end{equation}
hence, the right hand side above is {\it separable}: 
the first term 
depends on $(\nu)$ only while the second term, the summation,  on $(b)$ only.
Moreover,  
the second term is $\le 0$ provided
$(1+r_{\tiny \mbox{free}}) \delta_k \le 1$, 
along with $b$ being non-negative, part of the feasibility in (C2). 
In this case, we will have $U_k\le\mathbb{E} ( \Pi_{k,\tau_k}(\nu))$, 
which implies $U^*_k \le\max_{\tau_k, \nu} \mathbb{E}( \Pi_{k,\tau_k}(\nu))$,
with equality holding when $b_{k,t}=0$ for all $t=1, \dots, \tau_k$.

\quad We are ready to present the result on the utility maximisation problem in  \eqref{eq:OPT}. 
Two strategies are singled out:
the {\em ``buy-out'' strategy}, in which miner $k$ buys up all coins available at time $1$, and then 
participate in the PoS mining process until the end;
and the {\em ``non-participation'' strategy}, in which miner $k$ turns all $n_{k,0}$ coins into cash,
and then never participates in either PoS mining or trading for all $t\ge 1$.
Note that the non-participation strategy is executed at $\tau_k=0$; as such, it complements the feasible class, which is  
for $ \tau_k \ge 1$ and presumes participation.
The buy-out strategy clearly belongs to the feasible class.

\begin{theorem}[Buy-out strategy versus non-participation]
\cite{RS21, TYPoly}
\label{thm:4}
Assume the following two conditions: 
\begin{equation}
\label{condM}
{\rm (a)}\;\; \delta_k(1+r_{\tiny \mbox{free}}) \le 1 \quad{\rm and}\quad
{\rm (b)} \;\; \mathbb{E}(M_{t+1} \,|\, \mathcal{G}_t) = (1+r_{\tiny \mbox{cryp}}) M_t.
\end{equation}
Then with condition (a), the maximal utility
$U^*_k$ is achieved by setting $b_{k,t}=0$ for all $t=1, \dots, T_k$; i.e.,
$U^*_k =\max_\nu \mathbb{E}( \Pi_{k,T_k})$.
In addition, all three parts of the following will hold. 
\begin{enumerate}[itemsep = 3 pt]
\item[(i)] If
 $\delta_k  (1+r_{\tiny \mbox{cryp}}) \le 1$,
then 
any feasible strategy will provide no greater utility for miner $k$ than the non-participation strategy, 
 i.e., $U^*_k \le n_{k,0}P_0$.
\item[(ii)] If
$ \delta_k  (1+r_{\tiny \mbox{cryp}}) \ge 1$,
then any feasible strategy will provide no greater utility for miner $k$ than the buy-out strategy.
In this case, miner $k$ will buy all available coins at time $1$, and participate in the PoS mining process until the terminal time $T_k$.
\item[(iii)] If
 $\delta_k (1+r_{\tiny \mbox{cryp}})=1$, 
then miner $k$ is indifferent between
the non-participation and the buy-out strategy with any exit time, both of which will
provide no less utility than any feasible strategy. 
All strategies achieve the same utility (which is $\Pi_{k,0}=n_0P_{k,0}$).
\end{enumerate}
Moreover, when
$\delta_k =\delta:= (1+r_{\tiny \mbox{cryp}})^{-1}$ for all $k$, 
no miner will have any incentive to trade. 
Consequently, the long-term behavior of $\pi_{k,t}$ characterised in Theorems \ref{thm:1}--\ref{thm:3}
will hold.
\end{theorem}

\quad In what remains of this section, we make a few remarks on Theorem \ref{thm:4}, 
in particular, to motivate and explain its required conditions. 
First, the rate $r_{\tiny \mbox{cryp}}$, which is determined by condition (b),
is the (expected) rate of return of each coin, 
i.e. it is the counterpart of  $r_{\tiny \mbox{free}}$, the rate for the risk-free asset.
For all practical purpose, we can assume $r_{\tiny \mbox{crpt}}\ge r_{\tiny \mbox{free}}$, even though
this is not assumed in the theorem. 
When this relation holds, condition (a) will become
superfluous in cases (i) and (iii). 

\quad Second, the factor $\delta_k$ in the utility objective in \eqref{eq:OPT}, 
plays a key role in characterizing phase transitions in terms of $\delta_k(1+r_{\tiny{\mbox{cryp}}})$.
In case (i), the inequality $\delta_k\le 1/(1+r_{\tiny \mbox{cryp}})$ implies miner $k$ is seriously risk-averse; 
and this is reflected in $k$'s non-participation strategy.
In case (ii), the inequality holds in the opposite direction, implying miner $k$ is lightly risk-averse or even a risk taker.
Accordingly, $k$'s strategy is to aggressively sweep up all the available coins to reach monopoly, 
and participate (but not trade) until the terminal time.
In case (iii), the inequality becomes an equality $\delta_k = 1/(1+r_{\tiny \mbox{cryp}})$, and $(\Pi_{k,t})$ becomes a martingale.
Thus, miner $k$ is indifferent between non-participation and participation, 
and in the latter case, indifferent to all (feasible) strategies, including 
the buy-out (and the no-trading) strategy.
Indeed, the  equality $\delta_k = 1/(1+r_{\tiny \mbox{cryp}})$
is both necessary and sufficient for the no-trading strategy.

\quad Next, we emphasise that the two conditions in \eqref{condM} play very different roles.
Condition (b)
makes $(\Pi_{k,t})$ a super- or sub-martingale or a martingale,
according to miner $k$'s risk sensitivity as specified by the inequalities and equality applied to $\delta_k$
in the three cases.
Yet, to solve the maximisation problem in \eqref{eq:OPT}, 
$(\Pi_{k,t})$ needs to be connected to the utility; and this is the role played by
condition (a), under which it is necessary (for optimality) to set $b_{k,t}=0$ for all $t\ge 1$, 
and applicable to all three cases.
In this sense, condition (a) alone solves half of the maximisation problem, the $b_{k,t}$ half of the strategy.  
In fact, it's more than half, as the optimal $\nu$ strategy is only needed in the sub-martingale case; and even 
there, condition (a) pins down the fact that to participate (even without trading) is better than non-participation. 

\quad Theorem \ref{thm:4} is easily extended to the case where
the rates $r_{\tiny \mbox{cryp}}(t)$ and $r_{\tiny \mbox{free}}(t)$ may vary over the time. 
In this case, it suffices to modify the conditions in case (i) to 
$\left(1+ \sup_{t < T_k} r_{\tiny \mbox{cryp}}(t)\right) \delta_k \le 1$
and  
$\left(1+\sup_{t < T_k} r_{\tiny \mbox{free}}(t)\right) \delta_k \le 1$;
the conditions in case (ii) to
$\left(1+ \inf_{t < T_k} r_{\tiny \mbox{cryp}}(t)\right) \delta_k \ge 1$
and
$\left(1+\sup_{t < T_k} r_{\tiny \mbox{free}}(t)\right) \delta_k \le 1$;
and the conditions in case (iii) to 
$\delta_k = (1 + r_{\tiny \mbox{cryp}})^{-1}$
and 
$\sup_{t < T_k} r_{\tiny \mbox{free}}(t) \le r_{\tiny \mbox{cryp}}$.
Then, Theorem \ref{thm:4} will continue to hold. 

\quad Finally, the last part of the theorem considers the wealth evolution of a homogenous miner population.
It is also worth considering the wealth evolution of a heterogeneous miner population
(e.g. with different risk sensitivity, holding periods...etc.)
See \cite{JRS21} for a study on the reward effect on the wealth distribution 
of the miners with different coin holding horizons.

\section{PoS trading with volume constraint -- a continuous-time control setup}
\label{sc4}

\quad We continue to consider the question (2) in this section. 
As shown in Theorem \ref{thm:4}, 
the miner's strategy is either not to participate (in both PoS mining and trading),
or to sweep up all available coins immediately. 
The latter, being a market manipulation, rarely occurs in practice
due to regulation.
One way to prohibit the ``buy-out" strategy 
is to upper limit the volume that can be traded at a time.
This motivates the study of the PoS trading with volume constraint. 

\quad To simplify the analysis, 
we adopt a continuous-time control approach.
Time is continuous, indexed by $t\in [0,T]$,  for a fixed $T > 0$ representing the length of a finite horizon. 
Let $\{N(t), \, 0 \le t \le T\}$ (with $N(0):= N$) denote the process of the volume of coins,
which is increasing in time and sufficiently smooth.
So the derivative $N'(t)$ represents the instantaneous rate of ``reward'' by the PoS protocol.
For instance, we will consider below, as a special case, the process $N(t)$ of a polynomial form: 
\begin{equation}
\label{eq:Nal}
N_\alpha(t) = (N^{\frac{1}{\alpha}} + t)^\alpha, \qquad t \ge 0.
\end{equation}
The parametric family \eqref{eq:Nal} covers different rewarding schemes according to the values of $\alpha$:
for $0< \alpha < 1$, the process $N_{\alpha}(t)$ corresponds to a decreasing reward;
for $\alpha = 1$, the process $N_{1}(t) = N + t$ gives a rate one constant reward;
for $\alpha > 1$, the process $N_{\alpha}(t)$ amounts to an increasing reward.

\quad Let $K \ge 2$ be the number of miners, who are indexed by $k\in [K]:= \{1,\ldots, K\}$.
For each miner $k$, 
let $\{X_k(t), \, 0 \le t \le T\}$ (with $X_k(0) = x_k$) denote 
the process of the number of coins that miner $k$ holds,
with $X_k(t) \ge 0$ and $\sum_{k = 1}^K X_k(t) = N(t)$ for all $t\in [0,T]$.
For our continuous-time PoS model here, 
in which the time required for each round of voting is ``infinitesimal", 
imagine there are $M$ rounds of election during any given time interval $[t, t + \Delta t]$. 
(Each round in Ethereum takes about $10$ seconds, corresponding to the block-generation time \cite{BV14}.)
In each round miner $k$ gets either some coin(s) or nothing; 
so the average total number of coins $k$ will get over the $M$ rounds is
(by law of large numbers when $M$ is large),
\begin{equation*}
\underbrace{\frac{X_k(t)}{N(t)}  \frac{N'(t) \Delta t}{M}}_{\tiny \mbox{average number of coins in each round}} \times \underbrace{M}_{\tiny \mbox{number of rounds}} = \quad\frac{X_k(t)}{N(t)} N'(t) \Delta t.
\end{equation*}
Hence, replacing $\Delta t$ by the infinitesimal $dt$, 
we know miner $k$ will receive (on average) $\frac{X_k(t)}{N(t)}N'(t) dt$ coins, 
where $\frac{X_k(t)}{N(t)}$ is $k$'s winning probability, 
and $N'(t)  dt$ is the reward issued by the blockchain in $[t, t+dt]$.

\quad The miners are allowed to trade (buy or sell) their coins.
Miner $k$ will buy $\nu_k(t) dt$ coins in $[t, t+dt]$ if $\nu_k(t) > 0$,
and sell $-\nu_k(t) dt$ coins if $\nu_k(t) < 0$.
This leads to the following dynamics of miner $k$'s coins under trading:
\begin{equation}
\label{eq:Xnu}
X'_k(t) = \nu_{k}(t) + \frac{N'(t)}{N(t)} X_k(t) \quad \mbox{for } 0 \le t \le \tau_k  \wedge T:= \mathcal{T}_k,
\end{equation}
where $\tau_k: = \inf\{t>0: X_k(t) = 0\}$ is the first time at which the process $X_k(t)$ reaches zero.
It is reasonable to stop the trading process if a miner runs out of coins, or gets all available coins:
if $\mathcal{T}_k = \tau_k$, then miner $k$ liquidates all his coins by time $\tau_k$, and $X_k(\mathcal{T}_k) = 0$;
if $\mathcal{T}_k = \max_{j \ne k} \tau_j$, then miner $k$ gets all issued coins by time $\max_{j \ne k} \tau_j$,
and hence $X_k(\mathcal{T}_k) = N(\mathcal{T}_k)$.
We set $X_k(t) = X_k(\mathcal{T}_k)$ for $t > \mathcal{T}_k$.

\quad The problem is for each miner $k$ to decide how to trade coins with others under the PoS protocol.
Similar to Section \ref{sc3},
let $\{P(t), \, 0 \le t \le T\}$ be the price process of each (unit of) coin, 
which is a stochastic process assumed to be independent of the dynamics in \eqref{eq:Xnu}.
Let $b_{k}(t)$ denote the (units of) risk-free asset that miner $k$ holds at time $t$,
and let $r > 0$ denote the risk-free (interest) rate.
Recall that all $K$ miners are allowed to trade coins only internally among themselves,
whereas each miner can only exchange cash with an external source (say, a bank). 

\quad Let $\{c_k(t), \, 0 \le t \le T\}$ be the process of consumption, or cash flow of miner $k$, which follows the dynamics below: 
 \begin{equation}
dc_k(t) = rb_k(t) dt -db_k(t) - P(t) \nu_k(t) dt,  \qquad 
0 \le t \le \mathcal{T}_k; \tag{C1}
 \end{equation}
with 
 \begin{equation}
b_k(0)  = 0, \quad b_k(t) \ge 0 \mbox{ for } 0 \le t \le \mathcal{T}_k, \quad 0 \le X_k(t) \le N(t) \mbox{ for } 0 \le t \le \mathcal{T}_k. \tag{C2}
 \end{equation}
Set $b_k(t) = b_k(\mathcal{T}_k)$ and $\nu_k(t) = 0$ for $t > \mathcal{T}_k$.
The conditions (C1)--(C2) are the continuous analog 
to those in Section \ref{sc3}.
We also require that the trading strategy be bounded:
there is $\overline{\nu}_k > 0$ such that
\begin{equation}
| \nu_k(t) | \le \overline{\nu}_k. \tag{C3}
 \end{equation}
The objective of miner $k$ is:
\begin{equation}
\label{eq:obj1}
\begin{aligned}
\sup_{\{(\nu_k(t), b_k(t))\}} & J(\nu_k, b_k):=
\mathbb{E}\left\{ \int_0^{\mathcal{T}_k}e^{-\beta_k t} \left[dc_k(t) + \ell_k(X_k(t)) dt \right]  + e^{-\beta_k \mathcal{T}_k}  \left[b_k(\mathcal{T}_k) + h_k(X_k(\mathcal{T}_k) \right] \right\}  \\
& \mbox{ subject to } \eqref{eq:Xnu}, (\mbox{C}1), (\mbox{C}2), (\mbox{C}3),
\end{aligned}
\end{equation}
where $\beta_k > 0$ is a discount factor; 
$\ell_k(\cdot)$ and $h_k(\cdot)$ are two utility functions representing, respectively, the running profit and
the terminal profit.

\quad While generally following Merton's consumption-investment framework,
our formulation takes into account some distinct features of the PoS blockchain.
One notable point is, the utilities $\ell$ and $h$ are expressed 
as functions of the number of coins $X_k(t)$, as opposed to
their value $P(t) X_k(t)$. 
To the extent that $P(t)$ is treated as exogenous,
this difference may seem to be trivial.
Yet, it is a reflection of the more substantial fact that 
crypto-participants tend to mentally decouple the utility of holding coins from 
their monetary value at any given time.

\quad Throughout below, the following conditions will be assumed:

\smallskip
\begin{assump}~
\label{assump:1}
\begin{enumerate}[itemsep = 3 pt]
\item[(i)]
$N: [0,T] \to \mathbb{R}_{+}$ is increasing with $N(0) = N > 0$, and $N \in \mathcal{C}^2([0,T])$.
\item[(ii)]
$\ell: \mathbb{R}_{+} \to \mathbb{R}_{+}$ is increasing and $\ell \in \mathcal{C}^1(\mathbb{R}_+)$.
\item[(iii)]
$h: \mathbb{R}_{+} \to \mathbb{R}_{+}$ is increasing and $h \in \mathcal{C}^1(\mathbb{R}_+)$.
\end{enumerate}
\end{assump}

\quad To lighten notation, omit the subscript $k$, 
and write
\begin{align}
\label{eq:obj12}
U(x):= \sup_{\{(\nu(t), b(t))\}} & J(\nu,b):= \mathbb{E}\left\{ \int_0^{\mathcal{T}}e^{-\beta t} \left[dc(t) + \ \ell(X(t)) dt\right] + e^{-\beta \mathcal{T}} \left[ b(\mathcal{T}) + h(X(\mathcal{T}) \right]\right\} \\
& \mbox{ subject to } X'(t) = \nu(t) + \frac{N'(t)}{N(t)} X(t), \, X(0) = x, \tag{C0} \\
& \qquad \qquad \quad \, dc(t) =  rb(t)dt-db(t) - P(t) \nu(t) dt , \tag{C1} \\
& \qquad \qquad \quad \,  b(0) = 0, \, b(t) \ge 0 \mbox{ and } 0 \le X(t) \le N(t), \tag{C2} \\ 
& \qquad \qquad \quad  \, |\nu(t)| \le \overline{\nu}. \tag{C3}
\end{align}
Let
\begin{equation}
\label{eq:Pbeta}
\widetilde{P}_\beta(t): = e^{-\beta t} \mathbb{E} P(t) , \qquad  t\in[0, T].
\end{equation}
Substituting the constraint (C1) into the objective function, 
and taking into account $rb(t) dt -db(t)   = -e^{rt}d(e^{-rt} b(t))$,
along with \eqref{eq:Pbeta}, we have
\begin{equation}
\label{eq:42}
\begin{aligned}
J(\nu, b)
& = (r - \beta)\int_0^{\mathcal{T}} e^{-\beta t} b(t) dt + \int_0^\mathcal{T} \big[-\widetilde{P}_\beta(t) \nu(t) + e^{-\beta t}\ell(X(t)\big] dt + e^{-\beta \mathcal{T}}h(X(\mathcal{T})) \\
& :=J_1(b) + J_2( \nu).
\end{aligned}
\end{equation}
Hence,
\begin{equation}
\label{eq:U}
U(x) := \sup_{\{(\nu, b)\}} J( \nu,b) = \sup_{b} J_1(b) + 
\sup_{\nu} J_2(\nu).
\end{equation}

\quad Suppose $\beta\ge r$, which is analog to \eqref{condM}(a) in the discrete setting.
Then, from the $J_1(b)$ expression in \eqref{eq:42}, and taking into account $b(t) \ge 0$ as constrained in (C2), we have 
$\sup_b J_1(b) = 0$ with the optimality binding at $b_*(t) = 0$ for all $t$.
Therefore, 
the problem in \eqref{eq:obj12} reduces to
\begin{equation}
\label{eq:45}
U(x) = \sup_{\nu} J_2(\nu) \quad \mbox{subject to (C0), (C2'), (C3)},
\end{equation}
where (C2') is (C2) without the constraints on $b(\cdot)$. 
The problem \eqref{eq:obj12} can then be solved by 
dynamic programming and the Hamilton-Jacobi-Bellman (HJB) equations. 
The result is stated as follows. 

\begin{theorem}
\cite{TY23}
\label{thm:5}
Assume that $r \le \beta$, and $\widetilde{P}_{\beta}(t)$ in \eqref{eq:Pbeta} satisfies the Lipschitz condition:
\begin{equation}
\label{eq:LipP}
|\widetilde{P}_{\beta}(t) - \widetilde{P}_{\beta}(s)| \le C|t-s| \quad \mbox{for some } C >0.
\end{equation}
Then, $U(x) = v(0,x)$ where $v(t,x)$ is the unique viscosity solution to the following HJB equation,
where $Q: = \{(t,x): 0 \le t < T, \, 0<x<N(t)\}$:
\begin{equation}
\label{eq:HJB4}
\left\{ \begin{array}{lcl}
\partial_t v + e^{-\beta t} \ell(x) + \frac{x N'(t)}{N(t)} \partial_x v + \sup_{|\nu| \le \overline{\nu}} \{\nu ( \partial_x v - \widetilde{P}_{\beta}(t))\} = 0 \quad \mbox{in } Q , \\
v(T,x) = e^{-\beta T} h(x), \\
v(t,0) = e^{-\beta t} h(0), \,\, v(t, N(t)) = e^{-\beta t} h(N(t)).
\end{array}\right.
\end{equation}
Moreover, the optimal strategy is $b_{*}(t) = 0$ and $\nu_{*}(t) = \nu_{*}(t, X_{*}(t))$ for $0 \le t \le \mathcal{T}_{*}$,
where $\nu_{*}(t,x)$ achieves the supremum in \eqref{eq:HJB4}, and $X_{*}(t)$ solves 
$X_*'(t) = \nu_*(t, X_*(t)) + \frac{N'(t)}{N(t)}X_*(t)$ with $X_*(0) = x$, and $\mathcal{T}_{*}: = \inf\{t>0: X_*(t) = 0 \mbox{ or } N(t)\} \wedge T$.
\end{theorem}

\quad Now specialise to linear utility $\ell(x) = \ell x$ and $h(x) = hx$, for some 
given (positive) constants $\ell$ and $h$. 
In this case we can derive a closed-form solution to the HJB equation in \eqref{eq:HJB4}, 
and then derive the optimal strategy $\nu_*(t)$ (in terms of $\widetilde{P}_\beta(t)$).
Let 
\begin{equation}
\label{eq:410}
\Psi(t):=\frac{1}{N(t)} \left(h e^{-\beta T} N(T) + \ell \int_t^T  e^{-\beta s} N(s) ds \right).
\end{equation}
The following corollary classifies all possible optimal strategies (of the miner).

\begin{corollary}
\cite{TY23}
\label{coro:class}
Let $\ell(x) = \ell x$ and $h(x) = h x$ with $\ell, h > 0$, and $N(t)$ satisfy Assumption \ref{assump:1} (i).
Assume that $\widetilde{P}_\beta(t)$ satisfies the Lipschitz condition in \eqref{eq:LipP}, and that $ \overline{\nu}$ satisfies:
\begin{equation}
\label{eq:toend}
\overline{\nu} \int_0^T \frac{dt}{N(t)} \le \frac{x}{N} \wedge \frac{N-x}{N}.
\end{equation}
Then, the following results hold:
\begin{itemize}[itemsep = 3 pt]
\item[(i)]
Suppose $\widetilde{P}_\beta(t)$ stays constant, i.e., for all $t\in [0,T]$, 
 $\widetilde{P}(t) =\widetilde{P}(0)= P(0)$. 
\begin{enumerate}[itemsep = 3 pt]
\item[(a)]
If $P(0) \ge \Psi(0)$, 
then $\nu_*(t) = - \overline{\nu}$ for all $0 \le t \le T$. 
\item[(b)]
If $P(0) \le \Psi(T)$, then $\nu_*(t) = \overline{\nu}$.
\item[(c)]
If $\Psi(T) < P(0) < \Psi(0)$,
then 
$\nu_*(t)  = \overline{\nu}$ for $t \le t_0$, and $- \overline{\nu}$ for $t > t_0$,
where $t_0$ is the unique point in $[0,T]$ such that $P(0) =\Psi(t_0)$ with $\Psi(t)$ defined in \eqref{eq:410}.
\end{enumerate}
\item[(ii)]
Suppose that $\widetilde{P}_\beta(t)$ is increasing in $t\in [0,T]$.
\begin{enumerate}[itemsep = 3 pt]
\item[(a)]
If $P(0) \ge \Psi(0)$, 
then $\nu_*(t) = - \overline{\nu}$ for all $0 \le t \le T$. 
\item[(b)]
If $\widetilde{P}_\beta(T) \le \Psi(T)$, then $\nu_*(t) = \overline{\nu}$.
\item[(c)]
If $P(0) < \Psi(0)$ and $\widetilde{P}_\beta(T) > \Psi(T)$,
then 
$\nu_*(t) = \overline{\nu}$ for $t \le t_0$, and $-\overline{\nu}$ for $t > t_0$,
where $t_0$ is the unique point of intersection of $\widetilde{P}_\beta(t)$ 
and $\Psi(t)$ on $[0,T]$.
\end{enumerate}
\item[(iii)]
Suppose that $\widetilde{P}_\beta(t)$ is decreasing  in $t\in [0,T]$.
\begin{enumerate}[itemsep = 3 pt]
\item[(a)]
If $P(0) \ge \Psi(0)$, then the miner first sells, and may then buy, etc, always at full capacity,
according to the crossings of $\widetilde{P}_\beta(t)$ and $\Psi(t)$ in $[0,T]$.
\item[(b)]
If $P(0) < \Psi(0)$, then the miner first buys, and may then sell, etc, always at full capacity,
according to the crossings of $\widetilde{P}_\beta(t)$ and $\Psi(t)$ in $[0,T]$.
\end{enumerate}
\end{itemize}
\end{corollary}

\quad Several remarks are in order. 
First note that the condition in \eqref{eq:toend} is to guarantee the constraint (C2') not activated prior to $T$;
that is, to exclude the possibility of monopoly/dictatorship that will trigger a forced early exit.
Second, the monotone properties of $\widetilde{P}_\beta(t)$, which
classify the three parts (i)-(iii) in the corollary 
naturally connect to martingale pricing:
$\widetilde{P}_\beta(t)$ being a constant  in (i) makes the process $e^{-\beta t} P(t)$ a martingale;
whereas $\widetilde{P}_\beta(t)$ increasing or decreasing, respectively in (ii) and (iii),
makes $e^{-\beta t} P(t)$ a sub-martingale or a super-martingale.
On the other hand, the function $\Psi(t)$ 
represents the rate of return of the miner's utility (from holding of coins, $x$);
and interestingly, in the linear utility case, this return rate is independent of $x$ while decreasing in $t$.
Thus, the trading strategy is completely determined by comparing this return rate $\Psi(t)$ with the miner's
risk-adjusted coin price $\widetilde{P}_\beta(t)$:
if $\Psi(t)\ge ({\rm resp.} <) \widetilde{P}_\beta(t)$, then the miner will buy (resp.\ sell) coins.

\begin{figure}[htb]
    \centering
\begin{subfigure}{0.45\textwidth}
  \includegraphics[width=\linewidth]{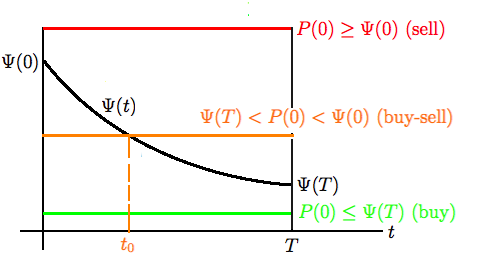}
\end{subfigure}\hfil
\begin{subfigure}{0.45\textwidth}
  \includegraphics[width=\linewidth]{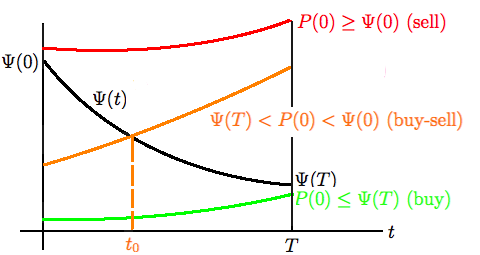}
\end{subfigure}\hfil
\caption{Optimal trading with linear $\ell(\cdot), h(\cdot)$ when $\widetilde{P}_\beta(t)$ is constant (left)
and $\widetilde{P}_\beta(t)$ is increasing (right).}
\label{fig:OT}
\end{figure}

\quad Specifically, following (i) and (ii) of Corollary \ref{coro:class},
for a constant or an increasing $\widetilde{P}_\beta(t)$
(corresponding to a risk-neutral or risk-seeking miner),
there are only three possible optimal strategies:
buy all the time, sell all the time, or first buy then sell.
(The first-buy-then-sell strategy echoes the general investment practice that
an early investment pays off in a later day.) 
See Figure \ref{fig:OT} for an illustration.

\quad In part (iii) of Corollary \ref{coro:class}, when $\widetilde{P}_\beta(t)$ is decreasing in $t$, like $\Psi (t)$,
the multiple crossings between the two decreasing functions can be further pinned down when there's more model structure.
Consider, for instance, when $P(t)$ follows a geometric Brownian motion (GBM):
\begin{equation}
\label{eq:gbm}
\frac{dP(t)}{P(t)}=\mu dt +\sigma dB_t, \quad{\rm or}\quad
P(t) = P(0) e^{(\mu - \sigma^2/2) t + \sigma B_t} ;  \quad t \in[0, T],
\end{equation}
where $\{B_t\}$ denotes the standard Brownian motion; and
$\mu> 0$ and  $\sigma > 0$ are the two parameters of the GBM model, representing the rate
of return and the volatility of $\{P(t)\}$.
The following proposition gives the conditions under which $\Psi_\alpha(t) - \widetilde{P}_\beta(t)$ is monotone
in the regime $N \to \infty$,
and optimal strategies are derived accordingly. 

\begin{figure}[h]
\centering
\includegraphics[width=0.45\columnwidth]{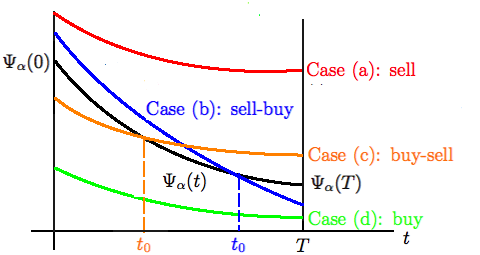}
\caption{Optimal trading with linear $\ell(\cdot), h(\cdot)$ when $\widetilde{P}_\beta(t)=P(0) e^{(\mu - \beta)t}$ and $N(t) = N_\alpha(t)$.}
\label{fig:OT2}
\end{figure}

\begin{proposition}
\cite{TY23}
Suppose the assumptions in Proposition \ref{coro:class} hold, 
with $N(t) = N_\alpha(t)$ and $\{P (t)\}$ specified by \eqref{eq:gbm} with $\beta > \mu$.
As $N \to \infty$, we have the following results: 
\begin{enumerate}[itemsep = 3 pt]
\item
If for some $\varepsilon > 0$,
$P(0) > \frac{1}{\beta - \mu} \left(\frac{\alpha h e^{-\mu T} (N^{\frac{1}{\alpha}} +T)^{\alpha}}{N^{1+ \frac{1}{\alpha}}} + \frac{\alpha \ell \beta^{-1}}{N^{\frac{1}{\alpha}}} + \ell \right) + \frac{\varepsilon}{N^{\frac{1}{\alpha}}}$,
then $\Psi_\alpha(t) - \widetilde{P}_\beta(t)$ is increasing on $[0,T]$.
\item
If for some $\varepsilon > 0$,
$P(0) < \frac{1}{\beta - \mu}\left(\frac{\alpha h e^{-\beta T}}{N^{\frac{1}{\alpha}}+T} + \ell e^{-\mu T} \right) - \frac{\varepsilon}{N^{\frac{1}{\alpha}}}$,
then 
$\Psi_\alpha(t) - \widetilde{P}_\beta(t)$ is decreasing on $[0,T]$.
\end{enumerate}
Consequently, we have:
\begin{enumerate}[itemsep = 3 pt]
\item[(a)]
If $P(0) > e^{(\beta - \mu)T} \Psi_\alpha(T)$ and (1) holds,
or $P(0) > \Psi_\alpha(0)$ and (2) holds,
then $\nu_*(t) = - \overline{\nu}$ for all $t$.
\item[(b)]
If $\Psi_\alpha(0) \le P(0) < e^{(\beta -\mu)T} \Psi_\alpha(T)$ and (1) holds,
then $\nu_*(t) = - \overline{\nu}$ for $t \le t_0$ and $\nu_*(t) =\overline{\nu}$ for $t > t_0$,
where $t_0$ is the unique point of intersection of $\widetilde{P}_\beta(t)$ 
and $\Psi_\alpha (t)$ on $[0,T]$.
\item[(c)]
If $e^{(\beta - \mu)T} \Psi_\alpha(T) \le P(0) < \Psi_\alpha(0)$ and (2) holds,
then $\nu_*(t) = \overline{\nu}$ for $t \le t_0$ and $\nu_*(t) = -\overline{\nu}$ for $t > t_0$,
where $t_0$ is the unique point of intersection of $\widetilde{P}_\beta(t)$ 
and $\Psi_\alpha (t)$ on $[0,T]$.
\item[(d)]
If $P(0) < e^{(\beta - \mu)T} \Psi_\alpha(T)$ and (2) holds,
or $P(0) < \Psi_\alpha(0)$ and (1) holds,
then $\nu_*(t) = \nu$ for all $t$.
\end{enumerate}
\end{proposition}
See Figure \ref{fig:OT2} for an illustration of all four possible strategies.

\section{PoS trading -- a mean field model}
\label{sc5}

\quad We consider the question (3) in this section. 
In the previous sections,
we study the optimal strategy for each individual miner,
assuming that all other miners will  ``cooperate" with her
(except that no shorting is allowed).
This assumption seems to be too optimistic.
As mentioned in part (c) in the introduction, 
there are also investors or speculators participating in the PoS trading. 
\begin{itemize}[itemsep = 3 pt]
\item
In Sections \ref{sc3}--\ref{sc4}, the price process (of each coin) is assumed to be exogenous. 
In the presence of investors and 
in view of their speculative nature, 
it is necessary to incorporate the market impact into the price formation.
\item
Interaction and competition among the miners and investors 
should play a role in each miner's decision. 
Thus, it is natural to formulate the PoS trading as a game 
building on the continuous-time control setting in Section \ref{sc4}.
\end{itemize}
With an exchange platform (e.g. Coinbase) for miner-investor tradings, 
we can define a notion of {\em equilibrium trading strategy} for a typical miner. 
This leads to the wealth evolution of the whole (miner) population from a mean field perspective.
Refer to \cite{BBLL20, LRS19, PW21} for discussions 
on the game theoretical analysis of the PoW protocol.
We would like to point out that
the material in this section is preliminary (and novel),
so there are many research problems in modeling, theory and applications.

\quad Recall that $\{N(t), \, 0 \le t \le T\}$ is the process of the volume of coins issued by
the PoS blockchain. 
There are $K$ miners, indexed by $k \in [K]$.
For each miner $k$, 
$\{X_k(t), \, 0 \le t \le T\}$ (with $X_k(0) = x_k$) denotes 
the process of the number of coins that miner $k$ holds,
and $\{\nu_k(t), \, 0 \le t \le T\}$ denotes miner $k$'s trading strategy.

\quad Let $\{Z(t), \, t \ge 0\}$ (with $Z(0) = z$) be the process 
of the number of coins that investors possess,
with $X_k(t)$, $Z(t) \ge 0$ and 
\begin{equation}
\label{eq:conserv}
\sum_{k = 1}^K X_k(t) + Z(t) = N(t),  \quad \mbox{for } 0 \le t \le T.
\end{equation}
So there are only $N(t) - Z(t)$ coins committed in the PoS election. 
The dynamics of miner $k$'s coin under trading is:
\begin{equation}
\label{eq:Xnu2}
X'_k(t) = \nu_{k}(t) + \frac{N'(t)}{N(t) - Z(t)} X_k(t) \quad \mbox{for } 0 \le t \le \tau_k  \wedge T:= \mathcal{T}_k,
\end{equation}
where $\tau_k: = \inf\{t>0: X_k(t) = 0\}$.
We set $X_k(t) = X_k(\mathcal{T}_k)$ for $t > \mathcal{T}_k$.
If we take $Z(t) \equiv 0$ (no investors), 
the dynamics \eqref{eq:Xnu2} reduces to \eqref{eq:Xnu}.
The equations \eqref{eq:conserv}--\eqref{eq:Xnu2} imply that $\sum_{k = 1}^K \nu_k(t) + Z'(t) = 0$,
which can be viewed as a clearing house condition.
It simply yields
\begin{equation}
\label{eq:Z}
Z(t) = Z(0) - \int_0^t \sum_{k = 1}^K \nu_k(s) ds.
\end{equation}

\quad Central to each miner's decision is the price process 
$\{P(t), \, t \ge 0\}$ of each (unit) of coin. 
The modern trading theory postulates a market impact structure underlying the price.
That is, the asset price is affected by the trading volume. 
For ease of presentation, we adopt the linear price impact, i.e. the Almgren-Chriss model \cite{AC01}:
\begin{equation}
\label{eq:ACP}
P(t) = P(0) + \sigma B(t) - \eta(Z(t) - Z(0)),
\end{equation}
where $\{B(t), \, t \ge 0\}$ is the standard Brownian motion,
$\sigma >0$ is the volatility, and $\eta > 0$ is the market impact parameter.
Refer to \cite[Chapter 3]{Gueant16} for background,
and \cite{DB15, GS11, TL11} for other market impact models.

\quad Recall that $b_{k}(t)$ is the (units of) risk-free asset that miner $k$ holds at time $t$,
and $r$ is the risk-free rate.
Here, all $K$ miners and investors can trade coins on the exchange platform,
whereas each miner can only exchange cash with an external source.
For each miner $k$, the process of consumption $\{c_k(t), \, 0 \le t \le T\}$
evolves as
\begin{equation}
dc_k(t) = rb_k(t) dt -db_k(t) - P(t) \nu_k(t) dt - N'(t) \, L\left(\frac{\nu_k(t)}{N'(t)}\right) dt,  \qquad 
0 \le t \le \mathcal{T}_k, \tag{C1}
\end{equation}
where $L(\cdot)$ is an even function, increasing on $\mathbb{R}_+$, strictly convex and asymptotically super-linear.
Compared to (C1) in Section \ref{sc4}, 
the additional term $- N'(t) \, L\left(\frac{\nu_k(t)}{N'(t)}\right)$ stands for the transaction cost 
which depends not only on the traded volume $\nu_k(t) dt$ 
but also the total volume $N'(t) dt$ (see \cite[p.43, (3.3)]{Gueant16}).
The quadratic cost $L(x) = \rho |x|^2$ with $\rho > 0$ corresponds to the original Almgren-Chriss model,
which we will mostly stick to.
We also impose the no shorting constraint:
 \begin{equation}
b_k(0)  = 0, \quad b_k(t) \ge 0 \mbox{ for } 0 \le t \le \mathcal{T}_k, \quad 0 \le X_k(t) \le N(t) \mbox{ for } 0 \le t \le \mathcal{T}_k. \tag{C2}
 \end{equation}
Set $b_k(t) = b_k(\mathcal{T}_k)$ and $\nu_k(t) = 0$ for $t > \mathcal{T}_k$.

\smallskip
{\bf Miner's strategy if $Z(\cdot)$ is known}.
It is easy to see from \eqref{eq:ACP} that 
the price $P(t)$ (up to noise) only depends on the investor holdings $Z(t)$,
or equivalently all the miners' holdings $\sum_{k =1}^K X_k(t)$.
Here, suppose that each miner $k$ ``knows" the number of coins that the investors hold. 
As in \eqref{eq:obj1}, the objective of miner $k$ is:
\begin{equation}
\label{eq:obj11}
\begin{aligned}
\sup_{\{(\nu_k(t), b_k(t))\}} & J(\nu_k, b_k):=
\mathbb{E}\left\{ \int_0^{\mathcal{T}_k}e^{-\beta_k t} \left[dc_k(t) + \ell_k(X_k(t)) dt \right]  + e^{-\beta_k \mathcal{T}_k}  \left[b_k(\mathcal{T}_k) + h_k(X_k(\mathcal{T}_k) \right] \right\}  \\
& \mbox{ subject to } \eqref{eq:Xnu2}, \eqref{eq:ACP}, (\mbox{C}1), (\mbox{C}2),
\end{aligned}
\end{equation}
Assuming that all the miners are {\em interchangeable} (which assumes a homogenous miner population),
we drop the subscript `$k$' in \eqref{eq:obj11},
and the objective of a {\em typical} miner is:
 \begin{align}
\label{eq:obj121}
U(x):= & \sup_{\{(\nu(t), b(t))\}} J(\nu,b):= \mathbb{E}\left\{ \int_0^{\mathcal{T}}e^{-\beta t} \left[dc(t) + \ \ell(X(t)) dt\right] + e^{-\beta \mathcal{T}} \left[ b(\mathcal{T}) + h(X(\mathcal{T}) \right]\right\} \\
& \mbox{ subject to } X'(t) = \nu(t) + \frac{N'(t)}{N(t) - Z(t)} X(t), \, X(0) = x, \tag{C0} \\
& \qquad \qquad \quad \, dc(t) =  rb(t)dt-db(t) - P(t) \nu(t) dt - N'(t) \, L\left(\frac{\nu(t)}{N'(t)} \right) dt, \tag{C1} \\
& \qquad \qquad \quad \,  b(0) = 0, \, b(t) \ge 0 \mbox{ and } 0 \le X(t) \le N(t), \tag{C2} \\
& \qquad \qquad \quad \, P(t) = P(0) + \sigma B(t) - \eta(Z(t) - Z(0)), \tag{C3}
\end{align}
where (C0) is a repeat of the state dynamics in \eqref{eq:Xnu2}, and (C3) is the price dynamics in \eqref{eq:ACP}.
Compared to \eqref{eq:obj12}, 
the volume constraint $|\nu(t)| \le \overline{\nu}$ for $\overline{\nu} > 0$ is removed;
instead the transaction cost $- N'(t) \, L\left(\frac{\nu(t)}{N'(t)} \right) dt$
is introduced in the budget constraint (C1).
This way, the miner's strategy will no longer be a bang-bang control 
but depend on the specific market impact mechanism.

\quad Let
\begin{equation}
\label{eq:Pbeta}
\widetilde{P}_\beta(t): = e^{-\beta t} \mathbb{E} P(t) = e^{-\beta t} \left[ P(0) - \eta (Z(t) - Z(0)) \right] \quad \mbox{and} \quad
\widetilde{P}(t): = \widetilde{P}_0(t).
\end{equation}
The same argument as in \eqref{eq:42}--\eqref{eq:U}
shows that the consumption-investment problem \eqref{eq:obj121} is separable:
\begin{equation*}
U(x) := \sup_{\{(\nu, b)\}} J( \nu,b) = \sup_{b} J_1(b) + 
\sup_{\nu} J_2(\nu),
\end{equation*}
where $J_1(b):=(r - \beta)\int_0^{\mathcal{T}} e^{-\beta t} b(t) dt$ and
\begin{equation*}
J_2(\nu):=\int_0^\mathcal{T} \left[-\widetilde{P}_\beta(t) \nu(t) - e^{-\beta t} N'(t) L \left( \frac{\nu(t)}{N'(t)}\right) + e^{-\beta t}\ell(X(t)\right] dt 
+ e^{-\beta \mathcal{T}}h(X(\mathcal{T})).
\end{equation*}

\quad Again suppose $\beta \ge r$. 
Then $\sup_b J_1(b) = 0$ with the optimality binding at $b_*(t) = 0$ for all $t$. 
So the problem \eqref{eq:obj12} is reduced to
\begin{equation}
\label{eq:45}
U(x) = \sup_{\nu} J_2(\nu) \quad \mbox{subject to (C0), (C2')},
\end{equation}
where (C2') is (C2) without the constraints on $b(\cdot)$. 

\quad Next we argue by dynamic programming, and let
\begin{align*}
v(t,x):= & \sup_{\{\nu(s), s \ge t\}} \int_t^{\mathcal{T}}
\left[-\widetilde{P}_\beta(s) \nu(s) - e^{-\beta s} N'(s) L \left( \frac{\nu(s)}{N'(s)}\right) + e^{-\beta s}\ell(X(s)\right] dt 
+ e^{-\beta \mathcal{T}}h(X(\mathcal{T})) \\
& \mbox{ subject to } X'(s) = \nu(s) + \frac{N'(s)}{N(s) - Z(s)} X(s), \, X(t) = x,  \\
& \qquad \qquad \quad \, 0 \le X(s) \le N(s),
\end{align*}
so $U(x) = v(0,x)$. 
Let $Q: = \{(t,x): 0 \le t < T, \, 0<x<N(t)\}$.
Under some suitable conditions, 
$v$ is the unique viscosity solution to the HJB equation:
\begin{equation*}
\left\{ \begin{array}{lcl}
\partial_t v + e^{-\beta t} \ell(x) + \frac{x N'(t)}{N(t)} \partial_x v + \sup_{\nu} \left\{\nu ( \partial_x v - \widetilde{P}_{\beta}(t)) - e^{-\beta t} N'(t) L\left(\frac{\nu}{N'(t)} \right) \right\} = 0 \quad \mbox{in } Q , \\
v(T,x) = e^{-\beta T} h(x), \\
v(t,0) = e^{-\beta t} h(0), \,\, v(t, N(t)) = e^{-\beta t} h(N(t)).
\end{array}\right.
\end{equation*}
By optimizing $\nu \to \nu ( \partial_x v - \widetilde{P}_{\beta}(t)) - e^{-\beta t} N'(t) L\left(\frac{\nu}{N'(t)} \right)$, 
we get
$\nu_* = N'(t) (L')^{-1} \left(e^{\beta t}  \partial_x v - \widetilde{P}(t)\right)$.
This yields the following nonlinear PDE:
\begin{equation}
\label{eq:HJB}
\left\{ \begin{array}{lcl}
\partial_t v + e^{-\beta t} \ell(x) + \frac{x N'(t)}{N(t) - Z(t)} \partial_x v + e^{-\beta t} N'(t) \bigg\{(e^{\beta t}  \partial_x v - \widetilde{P}(t)) (L')^{-1}(e^{\beta t}  \partial_x v - \widetilde{P}(t)) \\
\qquad \qquad  \qquad \qquad \qquad \qquad \qquad \qquad \, - L\left( (L')^{-1}(e^{\beta t}  \partial_x v - \widetilde{P}(t))\right) \bigg\} = 0 \quad \mbox{in }Q, \\
v(T,x) = e^{-\beta T} h(x), \\
v(t,0) = e^{-\beta t} h(0), \,\, v(t, N(t)) = e^{-\beta t} h(N(t)).
\end{array}\right.
\end{equation}
When $L(x) = \rho x^2$, the PDE \eqref{eq:HJB} specialises to
\begin{equation}
\label{eq:HJBsqrt}
\left\{ \begin{array}{lcl}
\partial_t v + e^{-\beta t} \ell(x) + \frac{x N'(t)}{N(t) - Z(t)} \partial_x v + \frac{e^{\beta t} N'(t)}{4 \rho}(\partial_x v - \widetilde{P}_\beta(t))^2 = 0 \quad \mbox{in }Q, \\
v(T,x) = e^{-\beta T} h(x), \\
v(t,0) = e^{-\beta t} h(0), \,\, v(t, N(t)) = e^{-\beta t} h(N(t)),
\end{array}\right.
\end{equation}
with the optimal strategy $\nu_*(t,x) = \frac{N'(t)}{2 \rho}\left(e^{\beta t}  \partial_x v(t,x) - \widetilde{P}(t)\right)$.
To simplify the presentation, we assume the quadratic cost (and \eqref{eq:HJBsqrt}) from now on.

\smallskip
{\bf Mean field strategy}.
Assume that the distribution of the miners by their coin holdings
are approximated by $m_0(x) dx$ at time $t =0$.
The goal is to find an equilibrium trading strategy
 $\nu^{\tiny \mbox{eq}}(t \,|\, m_0)$, or simply $\nu^{\tiny \mbox{eq}}(t)$,
 which can be viewed as the averaged trading strategy among all the miners.

\quad Now let's describe the mean field model.
\begin{enumerate}[itemsep = 3 pt]
\item
Since there are $K$ miners, by \eqref{eq:Z}, the investors' equilibrium holdings are:
\begin{equation}
\label{eq:Zeqbis}
Z^{\tiny \mbox{eq}}(t) = Z(0) - K \int_0^t \nu^{\tiny \mbox{eq}}(s) ds,
\end{equation}
\item
Given $Z^{\tiny \mbox{eq}}(\cdot)$,
the miner's optimal strategy is
\begin{equation}
\label{eq:nueqbis}
\nu^{\tiny \mbox{eq}}_*(t,x) = \frac{N'(t)}{2 \rho} ( e^{\beta t}\partial_x v(t,x) - \widetilde{P}(t)),
\end{equation}
where $v$ is the solution to \eqref{eq:HJBsqrt} with $Z(t) = Z^{\tiny \mbox{eq}}(t)$.
\item
The feedback control of a (typical) miner is 
$X'(t) = \nu_*^{\tiny \mbox{eq}}(t,X(t))+ \frac{N'(t)}{N(t) - Z^{\tiny \mbox{eq}}(t)} X(t)$.
Hence, the density of the miners by their coin holdings solves the continuity equation:
\begin{equation}
\label{eq:FP}
\partial_t m + \partial_x \left( \left( \nu_*^{\tiny \mbox{eq}}(t,x) + \frac{x N'(t)}{N(t) - Z^{\tiny \mbox{eq}}(t) } \right) m  \right) = 0, \quad m(0,x) = m_0(x).
\end{equation}
\item
The equilibrium trading strategy $\nu^{\tiny \mbox{eq}}(t)$ satisfies the fixed point equation:
\begin{equation}
\label{eq:fixedpt}
\int \nu_*^{\tiny \mbox{eq}}(t,x) m(t,x) dx = \nu^{\tiny \mbox{eq}}(t).
\end{equation}
\end{enumerate}

\begin{figure}[h]
\centering
\includegraphics[width=0.3\columnwidth]{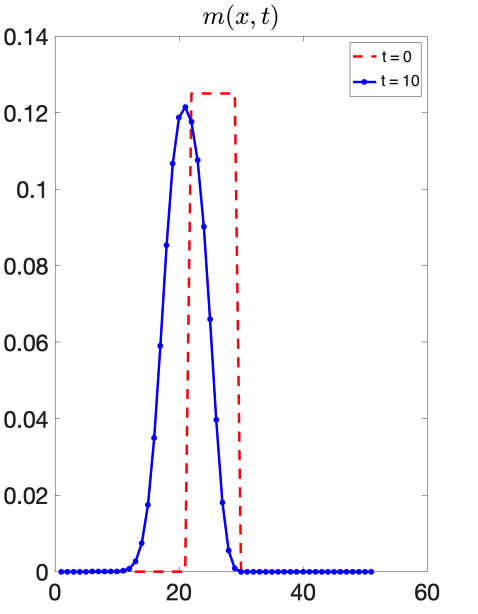}
\caption{Wealth evolution of the whole miner population}
\label{fig:WE}
\end{figure}

\quad If the mean field model \eqref{eq:Zeqbis}--\eqref{eq:fixedpt} is well-posed (i.e. has a unique solution),
then $m(t,\cdot)$ represents the wealth distribution of the whole miner population at time $t$.
To illustrate, 
Figure \ref{fig:WE} shows the wealth evolution of the miners
with initial shares uniformly distributed on $x \in [20, 30]$.
Observe that the wealth distribution spreads out and shifts to the left over the time,
which implies decentralisation of the PoS protocol.

\quad The problem now is to make rigorous such defined mean field model.
By injecting \eqref{eq:nueqbis} into \eqref{eq:fixedpt}, 
and then into \eqref{eq:Zeqbis}, \eqref{eq:HJBsqrt} and \eqref{eq:FP},
we get the mean field game form:
\begin{equation}
\label{eq:MFG}
\left\{ \begin{array}{lcl}
\partial_t v + H(t,x, \partial_x v, \partial_x v(\cdot, \cdot), m(\cdot, \cdot)) = 0,  \\
\partial_t m + \partial_p H(t,x, \partial_x v, \partial_x v(\cdot, \cdot), m(\cdot, \cdot)) = 0 = 0,  \\
m(t,0) = m_0(x), \quad  v(T, x) = e^{-\beta T} h(x),
\end{array}\right.
\end{equation}
with 
\begin{multline}
H(t,x,p, Q, m): =e^{-\beta t} \ell(x) +  \frac{2 \rho x N'(t)}{2 \rho(N(t) - Z(0)) + K \int_0^t N'(s) \int (e^{\beta s} Q(s,x) - \widetilde{P}(s) )m(s,x) dx ds} p  \\ + \frac{e^{\beta t} N'(t)}{4 \rho}(p - \widetilde{P}_\beta(t))^2,
\end{multline}
where
\begin{equation}
\label{eq:517}
\widetilde{P}_\beta(t) = e^{-\beta t} \left[ P(0) + \eta K \int_0^t \nu^{\tiny \mbox{eq}}(s) ds \right],
\end{equation}
and $\nu^{\tiny \mbox{eq}}(\cdot)$ satisfies
\begin{equation}
\label{eq:518}
\frac{N'(t)}{2 \rho} \left[e^{\beta t}  \int \partial_x v(t,x) m(t,x) dx - P(0) - \eta K \int_0^t \nu^{\tiny \mbox{eq}}(s)ds \right] 
= \nu^{\tiny \mbox{eq}}(t).
\end{equation}
Combining \eqref{eq:517} and \eqref{eq:518},
we have that $\widetilde{P}_\beta(t)$ (resp. $\widetilde{P}(t)$)
is a function of $Q$ and $m$ (i.e. $\partial_x v(\cdot, \cdot)$ and $m(\cdot, \cdot)$).

\quad The system of equations \eqref{eq:MFG} is a first-order mean field game 
with {\em non-local} and {\em non-separable} Hamiltonian.
It does not seem to have been studied before.
One possible idea to prove the wellposedness of \eqref{eq:MFG} is to
(1) add viscosity terms ($\varepsilon \Delta v$, $\varepsilon \Delta m$) to the equations,
and study the corresponding second-order mean field games (see \cite{Ambrose22});
(2) pass to the limit $\varepsilon \to 0$ by vanishing viscosity (see \cite{CG15, TZ23}).
The analysis of the equations \eqref{eq:MFG} 
may (probably) be very involved, 
and we leave it open.

\section{Conclusion}

\quad This chapter presents and surveys recent progress on
trading and wealth evolution in the PoS protocol
under various settings (discrete, continuous, volume constraint, transaction cost...etc.)
Below we provide a few open problems and future directions of research.
\begin{enumerate}[itemsep = 3 pt]
\item
In Section \ref{sc5}, we present a mean field model to study the wealth distribution
of the miners under the PoS protocol. 
This leads to a mean field game with non-local and non-separable Hamiltonian.
Prove that the equations \eqref{eq:MFG} have a (unique) solution.
\item
We have seen from Figure \ref{fig:WE} that the mean field model 
yields decentralisation.
Is it possible to prove a quantitative result to support this observation?
\item
In Sections \ref{sc3}--\ref{sc5}, we consider the optimal strategy of the miner assuming that 
$\delta_k (1 + r_{\tiny \mbox{free}}) \le 1 $ or $\beta \ge r$,
i.e. risk-averse. 
What's the miner's optimal strategy if she is more risk-seeking 
(such that $\delta_k (1 + r_{\tiny \mbox{free}}) > 1 $ or $\beta < r$)?
\item
We assume a fixed miner population, i.e. $K$ is fixed. 
In practice, some existing miners may quit, and some new miners may join at random times.
What happens if there is a dynamic miner population (or $K$ is varying over the time)?
\item
We assume that the reward $R_t$ is deterministic and is from the blockchain.
But in many PoS blockchains (e.g. Ethereum),
revenue for the miners comes mostly from the transaction fees. 
Thus, the ``reward" is from the user-miner interface (part (a)),
and the users can also be the investors. 
What will be the miner's strategy, and the wealth evolution
if we take the user-miner connection into account?
\item
In Sections \ref{sc3} and \ref{sc5}, we consider the wealth evolution of a homogenous population 
-- that is, all the miners solve the same optimisation problem.
What are the corresponding results 
for a heterogeneous miner population (e.g. with different risk sensitivity)?
\item
We assume that the miners maximise some objective to find the best strategy.
In practice, when people make decisions, 
they will adopt ``rational" strategy rather than the ``optimal" strategy.
This leads to the idea of {\em bounded rationality} \cite{Simon90},
which can be formulated in Bayesian languages. 
What is the miner's rational strategy?
\end{enumerate}

\bigskip
{\bf Acknowledgement:} 
We thank David Yao for collaborations which lead to a large part of the material presented in this chapter. 
We thank Erhan Bayraktar for pointing out the reference \cite{Ambrose22} on the mean field game with non-separable Hamiltonian.
We also thank Siyao Jiang and Yuhang Wu for help in numerical experiments. 
This research is supported by NSF grants DMS-2113779 and DMS-2206038,
and a start-up grant at Columbia University.

\bibliographystyle{abbrv}
\bibliography{unique}
\end{document}